\documentclass[smallextended]{svjour3} 
\usepackage{inputenc}
\expandafter\let\csname equation*\endcsname\relax
  \expandafter\let\csname endequation*\endcsname\relax
  \usepackage[colorlinks={true}]{hyperref}
\hypersetup{citecolor={blue}, filecolor={blue}, linkcolor={blue}, urlcolor={blue}}
\usepackage{mathtools}
\usepackage{graphics} 
\usepackage{epsfig} 
\usepackage{mathptmx} 
\usepackage{bm}
\usepackage{times,yhmath} 
\usepackage{amssymb, amsmath, mathrsfs, multicol}

\usepackage{amsthm}
\usepackage{cite}

\usepackage{thmtools}
\usepackage{etoolbox}
\usepackage{subeqnarray}
\usepackage{epstopdf} 

\DeclareMathOperator{\diag}{diag} %
\DeclareMathOperator{\tr}{tr} %
\DeclareMathOperator{\im}{Im} %
\DeclareMathOperator{\re}{Re} %
\DeclareMathAlphabet{\mathpzc}{OT1}{pzc}{m}{it}
\allowdisplaybreaks
\theoremstyle{definition}

\begin{document}
\title{Numerical detection of Gaussian entanglement and its application to the identification of bound entangled Gaussian states}

\author{Shan Ma\and Shibei Xue\and  Yu Guo\and Chuan-Cun~Shu}

\institute{S. Ma is with School of Automation, Central South University, Changsha 410083, China; He is also with Peng Cheng Laboratory, 518000, Shenzhen, China.  S. Xue is with Department of Automation, Shanghai Jiao Tong
University, Shanghai 200240, China; He is also with Key Laboratory of System Control and Information Processing,
Ministry of Education of China, Shanghai 200240, China. Y. Guo is with  Hunan Provincial Key Laboratory of Flexible Electronic Materials Genome Engineering, School of Physics and Electronic Science, Changsha University of Science and Technology, Changsha 410114, China. C. C. Shu is with Hunan Key Laboratory of Super-Microstructure and Ultrafast Process, School of Physics and Electronics, Central South University, Changsha 410083, China. \email{cc.shu@csu.edu.cn} }
\date{Received: date / Accepted: date}
\maketitle
\begin{abstract}
We present a numerical method for solving the separability problem of Gaussian quantum states in continuous-variable quantum systems.  We show that  the separability problem can be cast as an equivalent problem of determining the feasibility of a set of linear matrix inequalities. Thus, it can be efficiently solved  using existent numerical solvers. We apply this method to the identification of bound entangled Gaussian states. We show that  the proposed method can be used to identify bound entangled Gaussian states that could be simple enough to be producible in quantum optics.     
\end{abstract}

\section{Introduction}\label{sec:intro}
Quantum entanglement plays a central role in quantum information technologies, e.g., in quantum computation, quantum communication, and quantum metrology~\cite{HHHH09:rmp,BV05:rmp,AI07:jpa,weedbrook12:rmp,ARL14:OSID,NLKA18:pra,LYLW20:jpa}. In recent years,  a great deal of research effort has been put into the analysis of the entanglement properties of multiparticle systems~\cite{SAI05:pra,KCP07:pra,GK07:jpa,HV11:pra,MKJPCP12:jpa,LGZK13:jpb,SP18:ol,ODPACMWS18:nature,GLMS19:pra,GSDN19:prl}. While most of the effort has been devoted to systems with finite-dimensional Hilbert spaces, in particular discrete-variable qubit states, recently there has been considerable interest in the continuous-variable (CV) case~\cite{KY12:pra,C13:qip,
 JKN15:jpa,KLRA15:prl,GLS16:CP,GM16:pra,WGKWC04:pra,RR05:pra,WWLWW16:qip}. Gaussian states, as a particularly useful class of CV states, are commonly produced in quantum optics laboratories. Given a Gaussian state of a bipartite CV system, the most fundamental problem in CV quantum information theory is to determine whether the state is entangled or not  with respect to the splitting.  Consider two CV quantum systems $A$ with $m$ modes and $B$ with $n$ modes having infinite dimensional Hilbert spaces $\mathcal{H}_{A}$ and $\mathcal{H}_{B}$, respectively. The global bipartite system $A+B$ with $m+n$ modes has a Hilbert space $\mathcal{H}=\mathcal{H}_{A}\otimes\mathcal{H}_{B}$. By definition, a quantum state $\hat{\rho}$ of the global bipartite system $A+B$ is said to be separable if it can be written as a convex sum of pure product states, namely, 
\begin{align}
\hat{\rho}=\sum\limits_{j}p_{j}\hat{\rho}_{j}^{A}\otimes \hat{\rho}_{j}^{B}, \label{entangle_def}
\end{align}
where $p_{j}\ge 0$ and $\sum_{j}p_{j}=1$~\cite{W89:pra,weedbrook12:rmp}. Note that in Eq.~\eqref{entangle_def}, the sum can also be an integral and the probabilities are then replaced by a continuous probability density function.  Physically, Eq.~\eqref{entangle_def} means that separable states can be produced from product states by means of local  operations and classical communications (LOCCs). By definition, entangled states are states which are not separable.  Then the so-called separability problem is to determine whether a given quantum state is separable or not. Despite considerable progress made in recent years, the separability problem is still far from being completely solved~\cite{DGCZ00:prl,S00:prl,WW01:prl,GKLC01:prl,VF03:pra,HE06:njp,LSA18:njp}.

Perhaps the most commonly used tool in quantum information theory for checking if a given state is separable or not is based on the  partial transpose~\cite{HHH96:pla,P96:prl}. For a separable quantum state $\hat{\rho}$ as in Eq.~\eqref{entangle_def},  the partial transpose with respect to one of the two subsystems yields again a legitimate density operator and, in particular, positive, i.e., $\hat{\rho}^{T_{A}}=\sum_{j}p_{j}\left(\hat{\rho}_{j}^{A}\right)^{T}\otimes \hat{\rho}_{j}^{B}\ge 0$. Hence the positivity of the partial transpose (PPT) provides us a necessary condition for separability. However, it should be noted that the PPT criterion is, in general, not a sufficient condition for separability. In fact, a $2\times 2$-mode Guassian state, which has positive partial transpose but nevertheless is entangled, has been constructed in Ref.~\cite{WW01:prl}. This type of Gaussian  state is known as a bound entangled Gaussian state. Bound entangled Gaussian states are entangled states but their entanglement cannot be distilled into maximally entangled pure states with LOCCs~\cite{GDCZ01:QIC,Z11:pra,DSHPES11:prl,JZWZXP12:prl,SDSV14:pra,MWJZ19:pra}.  

For Gaussian quantum states, all the entanglement information is contained in the covariance matrix of position and momentum observables~\cite{WGKWC04:pra,weedbrook12:rmp}. Thus, the separability problem can be investigated at the level of covariance matrices. In fact, both the separability problem and the PPT criterion have been successfully reformulated in terms of the covariance matrix language in Ref.~\cite{WW01:prl}. Built upon this work, Ref.~\cite{GKLC01:prl} proposes a nonlinear iterative procedure to check the separability of a Gaussian state. Ref.~\cite{HE06:njp} developed a numerical method for finding an optimal entanglement witness that robustly detects the entangled state. Both methods  are nice and  effective for Gaussian states.  
 
In this paper, we  show that the separability problem can be cast as an equivalent problem of solving a set of linear matrix inequalities. If there exist solutions to the set of linear matrix inequalities, then the Gaussian state is separable; otherwise it is entangled. Thus the feasibility of the linear matrix inequalities serves as a necessary and sufficient condition for the separability of the corresponding Gaussian state.  On the other hand, solving linear matrix inequalities is a mature technology \cite{BGFB94:book}. There are many efficient numerical methods that can be used to solve linear matrix inequalities. For example, the linear matrix inequalities solver  SeDuMi~\cite{S99:oms} or  SDTP3 \cite{TTT03:mp} with YALMIP interface \cite{L04:conf} can perform very well. Thus, Gaussian entanglement can, in principle, be detected by using this method. Strictly speaking, the numerical method  proposed in this paper may be considered as a special case of the numerical method proposed in   Ref.~\cite{HE06:njp}. It receives particular attention in this paper since the method of solving linear matrix inequalities is  quite familiar to researchers from engineering~\cite{BGFB94:book}. By using the above LMI solvers, Gaussian entanglement can be detected very efficiently.    As an application, we use this numerical method to identify bound entangled Gaussian states. It is well known that bound entanglement is a rare phenomenon~\cite{HHHH09:rmp} and the detection of bound entanglement is a challenging problem~\cite{ZNZG10:pra}. In this paper, we use the proposed numerical method to identify a $2\times 2$ bound entangled Gaussian state that is easily  prepared in quantum optics.  We first parametrize the covariance matrices of $2\times 2$ Gaussian states using some free parameters. Then we choose the values of these parameters and see if the resulting covariance matrix passes the PPT and entanglement tests. If so, then we know the corresponding Gaussian state is a bound entangled Gaussian state. Based on the decomposition result, we can further provide a theoretical optical implementation for the generation of the bound entangled state. Finally, we highlight some important characteristics of this bound entangled state example, which may make it  the best candidate  for a realistic experimental verification of bound entanglement.
 
\section{Preliminaries}
We consider CV quantum systems with Gaussian states. Quantum continuous variables describe quantum mechanics applied to an infinite-dimensional Hilbert space equipped with a set of canonical quadrature operators $\hat{q}_{j}$ and $\hat{p}_{j}$ ($j=1,2,\cdots,n$). Here $\hat{q}_{j}$ and $\hat{p}_{j}$ are position and momentum operators, respectively. They  obey canonical commutation relations $[\hat{q}_{j},\hat{p}_{k}]=i\delta_{jk}$ (in natural units, $\hbar=1$).  The quadrature operators $\hat{q}_{j}$ and $\hat{p}_{j}$ are collected to form a vector of operators $\hat{\xi}=(\hat{q}_{1},\hat{p}_{1},\cdots,\hat{q}_{n},\hat{p}_{n})^{T}$. Then the  commutation relations  can be written as
\begin{align} 
[\hat{\xi}_{j} , \hat{\xi}_{k} ] = i\Omega_{jk},  \label{commutation}
\end{align}
where $\Omega_{jk}$ is the generic entry of the $2n \times 2n$  matrix $\Omega:=\omega^{\oplus n}=\begin{pmatrix}
\omega&&0\\
&\ddots\\
0&&\omega
\end{pmatrix}$, $\omega:=\begin{pmatrix}
0 &1\\
-1 &0
\end{pmatrix}$.
 
Gaussian states are CV states with Gaussian characteristic functions. Gaussian states are completely characterized by a real displacement vector $\langle \hat{\xi} \rangle=\tr (\hat{\rho}\hat{\xi})$ and a real  covariance matrix $\gamma$ with elements 
\begin{align}
\gamma_{jk}=\tr(\hat{\rho}\{\hat{\xi}_{j}-\langle\hat{\xi}_{j} \rangle , \hat{\xi}_{k}-\langle\hat{\xi}_{k} \rangle\}) \label{covariance}
\end{align} where we define the anticommutator product as $\{\hat{A}, \hat{B} \}=\hat{A}\hat{B}+\hat{B}\hat{A}$. The covariance matrix $\gamma$ is real and symmetric and due to the commutation rules \eqref{commutation}, the covariance matrix $\gamma$ satisfies the uncertainty relation 
\begin{align}
\gamma+i\Omega\ge 0. \label{uncertainty}
\end{align}
Inequality~\eqref{uncertainty} is a necessary and sufficient condition for a real symmetric matrix $\gamma$ to correspond to a physical quantum state~\cite{SSM87:pra}. 
The displacement vector $\langle \hat{\xi} \rangle$ contains no entanglement information. All the entanglement information is  contained in the covariance matrix of the state. Thus, we will restrict ourselves to the analysis of covariance matrices $\gamma$.  For a Gaussian state $\hat{\rho}_{AB}$ pertaining to an $(m+n)$-mode bipartite system $A+B$, a necessary and sufficient condition has been developed in Ref.~\cite{WW01:prl} for the separability of the state. It states that an $(m+n)$-mode  Gaussian state $\hat{\rho}_{AB}$   with  covariance matrix $\gamma$ is separable if and only if there exist an $m$-mode covariance matrix $\gamma_{A}\ge i \Omega_{A}$ and an $n$-mode covariance matrix $\gamma_{B}\ge i \Omega_{B}$ such that 
\begin{align}
\gamma\ge \gamma_{A}\oplus \gamma_{B}. \label{separability}
\end{align}
Here  $\gamma_{A}\oplus \gamma_{B}$ denotes the matrix direct sum of $\gamma_{A}$ and $\gamma_{B}$; that is, $\gamma_{A}\oplus \gamma_{B}=\begin{pmatrix}\gamma_{A} &0 \\ 0&\gamma_{B}\end{pmatrix}$. The condition~\eqref{separability} is a necessary and sufficient condition for the separability of a Gaussian state.  Given the covariance matrix $\gamma$, if we can find   $\gamma_{A}$ and $\gamma_{B}$  that satisfy the inequality~\eqref{separability}, then the state is separable; otherwise, it is entangled.  
 
A very convenient method for checking if a given state is separable or not is the PPT criterion. The PPT criterion provides a necessary condition for a state to be separable.  The partial transpose of a state   corresponds physically to a local time reversal. For a bipartite Gaussian system $A+B$, the partial transpose with respect to system $A$ transforms the  covariance matrix $\gamma$ into $\tilde{\gamma}=(\Lambda \oplus I_{B})\gamma (\Lambda \oplus I_{B})$, where $\Lambda=\oplus_{k=1}^{m}\diag(1,-1)$ corresponds to a sign change of the momentum variables belonging to system $A$ and $I_{B}$ is the $n$-mode identity matrix.  Clearly, if the partial transpose of $\hat{\rho}$ is a valid density operator, we must have $\tilde{\gamma}+i\Omega\ge 0$. But this is equivalent to $\gamma+i\tilde{\Omega}\ge 0$, where $\tilde{\Omega}=(-\Omega_{A})\oplus \Omega_{B}$.  Summarizing, suppose $\gamma$ is the covariance matrix of a Gaussian state, with finite second moments, which has positive partial transpose. Then we have 
\begin{align}
\gamma+i\tilde{\Omega}\ge 0,\; \text{where}\; \tilde{\Omega}=\begin{pmatrix}
-\Omega_{A} &0\\
0 &\Omega_{B}
\end{pmatrix}. \label{PPTcriterion}
\end{align}
The PPT criterion \eqref{PPTcriterion} is a necessary condition for separability. This can be seen from the fact that the inequality~\eqref{PPTcriterion} directly follows from the inequality~\eqref{separability}. In general, the PPT criterion is not sufficient to guarantee separability.  There exist Gaussian states that satisfy the PPT criterion but nevertheless are entangled. Such states are known as bound entangled Gaussian states. Bound entangled states are entangled states, but they cannot be distilled into pure entangled states using LOCCs. Bound entangled states have practical applications in quantum cryptography~\cite{AH09:pra}, channel discrimination~\cite{PW09:prl} and many quantum information protocols~\cite{JZWZXP12:prl}. 

\section{Detection of Gaussian Entanglement via Solving Linear Matrix Inequalities} \label{detection}
As mentioned in Eq.~\eqref{separability}, an $(m+n)$-mode Gaussian state $\hat{\rho}_{AB}$   with  covariance matrix $\gamma$ is separable if and only if there exist an $m$-mode covariance matrix $\gamma_{A}\ge i \Omega_{A}$ and an $n$-mode covariance matrix $\gamma_{B}\ge i \Omega_{B}$ such that 
$\gamma\ge \gamma_{A}\oplus \gamma_{B}$. Since the inequality $\gamma_{A}\ge i \Omega_{A}$ is equivalent to $\begin{pmatrix}
\gamma_{A} &\Omega_{A}\\
\Omega_{A}^{T} &\gamma_{A}
\end{pmatrix}\ge 0$ and the inequality $\gamma_{B}\ge i \Omega_{B}$ is equivalent to $\begin{pmatrix}
\gamma_{B} &\Omega_{B}\\
\Omega_{B}^{T} &\gamma_{B}
\end{pmatrix}\ge 0$ (see Lemma 2 in Ref.~\cite{GKLC01:prl} for a proof), the separability problem of a Gaussian state $\hat{\rho}_{AB}$ can be recast as an equivalent  feasibility problem involving linear matrix inequalities as follows: \\

\hrule\vspace{0.2cm}

\noindent\emph{Separability problem} 
\begin{align}
\textbf{find}\quad\; &\gamma_{A} \;\; \text{and}\;\; \gamma_{B} \notag  \\
\textbf{subject to}\quad\;  &\gamma- \begin{pmatrix}\gamma_{A} &0\\
                   0 &\gamma_{B}\end{pmatrix}\ge 0, \label{constraint1} \\
                   &\begin{pmatrix}
\gamma_{A} &\Omega_{A}\\
\Omega_{A}^{T} &\gamma_{A}
\end{pmatrix}\ge 0, \label{constraint2}  \\
&\begin{pmatrix}
\gamma_{B} &\Omega_{B}\\
\Omega_{B}^{T} &\gamma_{B}
\end{pmatrix}\ge 0. \label{constraint3} 
\end{align}  
\hrule \vspace{0.2cm}
The separability problem is thus to determine whether the constraints~\eqref{constraint1}, \eqref{constraint2} and \eqref{constraint3} are consistent for a given bipartite Gaussian state $\gamma$, and if so, find  a pair of positive definite matrices $\gamma_{A}$ and $\gamma_{B}$ satisfying  them. If the above problem is feasible, that is, there exist solutions  $\gamma_{A}$ and $\gamma_{B}$ to the constraints~\eqref{constraint1}, \eqref{constraint2}, and \eqref{constraint3}, then we can conclude that the covariance matrix $\gamma$ is separable; otherwise it is entangled. The linear matrix inequalities~\eqref{constraint1}, \eqref{constraint2} and \eqref{constraint3} can be  solved efficiently by using existent numerical solvers such as SeDuMi~\cite{S99:oms} and the SDTP3~\cite{TTT03:mp}. 

It should be mentioned that although the linear matrix inequalities~\eqref{constraint1}, \eqref{constraint2} and \eqref{constraint3} are necessary and sufficient for checking separability of $\gamma$,  one has to be very careful when the state $\gamma$ lies very close to the boundary of the set of separable states or  to the boundary of the set of  physical states (i.e., the smallest eigenvalue of $\gamma+i\Omega$ is very close to zero). For such cases, since the constraints~\eqref{constraint1}, \eqref{constraint2} and \eqref{constraint3} are non-strict linear matrix inequalities,  the unavoidable round-off errors caused by floating point computations may have a significant impact on the solvability of the problem. We consider an example to illustrate this fact. 

\emph{Example.}
Consider the Gaussian state constructed in Ref.~\cite{WW01:prl}. The covariance matrix of this $2\times 2$ Gaussian state is given by 
\begin{align}
\gamma=\begin{pmatrix}
2 &0 &0 &0 &1 &0 &0 &0\\
0 &1 &0 &0 &0 &0 &0 &-1\\
0 &0 &2 &0 &0 &0 &-1 &0\\
0 &0 &0 &1 &0 &-1 &0 &0\\
1 &0 &0 &0 &2 &0 &0 &0\\
0 &0 &0 &-1 &0 &4 &0 &0\\
0 &0 &-1 &0 &0 &0 &2 &0\\
0 &-1 &0 &0 &0 &0 &0 &4
\end{pmatrix}. \label{bound_state_example}
\end{align}
The eigenvalues of $\gamma+i\Omega$ are $0$, $3-\sqrt{3}$, $3$, $3+\sqrt{3}$, each with multiplicity $2$. Because $0$ is an eigenvalue of $\gamma+i\Omega$, the state lies just on the boundary of  the set of  physical states. For this particular $\gamma$, even if numerical computation shows that the problem with constraints~\eqref{constraint1}, \eqref{constraint2} and \eqref{constraint3} is infeasible, it is still too early to make a conclusion on whether the state is entangled or not, since we have not yet ruled out the possibility that the infeasibility is a result of the application of floating point computations to non-strict inequalities.  Fortunately, for this particular $\gamma$, we can relax the problem and solve the following strict inequalities instead:  \\
\begin{align}
\textbf{find}\quad\; &\gamma_{A} \;\; \text{and}\;\; \gamma_{B} \notag  \\
\textbf{subject to}\quad\;  &\gamma- \begin{pmatrix}\gamma_{A} &0\\
                   0 &\gamma_{B}\end{pmatrix}>-\epsilon I, \label{constraint4} \\
                   &\begin{pmatrix}
\gamma_{A} &\Omega_{A}\\
\Omega_{A}^{T} &\gamma_{A}
\end{pmatrix}> -\epsilon I, \label{constraint5}  \\
&\begin{pmatrix}
\gamma_{B} &\Omega_{B}\\
\Omega_{B}^{T} &\gamma_{B}
\end{pmatrix}>-\epsilon I. \label{constraint6} 
\end{align}
Here $\epsilon>0$ is a small number such that the LMI problem \eqref{constraint4}, \eqref{constraint5} and \eqref{constraint6}  \emph{slightly} relaxes the LMI problem \eqref{constraint1}, \eqref{constraint2} and \eqref{constraint3}.  Typically, we may choose $10^{-6}<\epsilon<10^{-9}$.  For any $\epsilon>0$,  if there is no solution $\gamma_{A}$, $\gamma_{B}$ to the inequalities~\eqref{constraint4}, \eqref{constraint5} and \eqref{constraint6},  then there is no solution $\gamma_{A}$, $\gamma_{B}$ to the LMI problem \eqref{constraint1}, \eqref{constraint2} and \eqref{constraint3}, and we can safely conclude that the state $\gamma$ is an entangled state. For this particular example, we choose $\epsilon=10^{-8}$. We find that even in this case, there is still no solution $\gamma_{A}$, $\gamma_{B}$ to the relaxed inequalities~\eqref{constraint4}, \eqref{constraint5} and \eqref{constraint6}. So we can safely conclude that the state $\gamma$ in Eq.~\eqref{bound_state_example} is an entangled state. On the other hand, a direct  calculation shows that $\gamma$ has a positive partial transpose, i.e., $\gamma+i\tilde{\Omega}\ge 0$. Thus the state with covariance matrix~\eqref{bound_state_example} is a bound entangled Gaussian state with respect to the bipartite splitting  $\{\{1,\; 2\},\;\{3,\; 4\}\}$.   

\begin{remark}
We make some remarks about how to solve linear matrix inequalities numerically. Linear matrix inequalities are essentially convex constraints and can be solved efficiently using many existing numerical methods~\cite{BGFB94:book,BV04:book}. A simple algorithm that is guaranteed to solve linear matrix inequality problems is the ellipsoid algorithm~\cite{BBT81:or}. A more computationally efficient algorithm is the interior point method~\cite{NN94:book}. The basic idea of these algorithms can be found in~\cite{BGFB94:book}. Based on these algorithms, several software packages for solving linear matrix inequalities have been produced such as the solvers SeDuMi~\cite{S99:oms} and the SDTP3~\cite{TTT03:mp}. We refer the reader to the appendices of Ref.~\cite{ZSN16:book} for some code examples on how to use these solvers. 
\end{remark} 

\section{Identification of $2\times 2$ bound entangled Gaussian states}
Bound entangled Gaussian states are a class of Gaussian states that satisfy the PPT criterion, but nevertheless are entangled. In order to guarantee  a bound entangled state, we need to make sure that the following two conditions hold: 1) the covariance matrix $\gamma$ satisfies the PPT criterion \eqref{PPTcriterion}; 2) the linear matrix inequalities~\eqref{constraint1}~-~\eqref{constraint3} are infeasible; that is, there exist no solutions $\gamma_{A}$ and $\gamma_{B}$ to the constraints~\eqref{constraint1}~-~\eqref{constraint3}. 

According to a theorem by Williamson~\cite{W36:ajm}, every positive-definite real symmetric matrix of even dimension can be diagonalized through a symplectic transformation. In particular, this theorem can be applied to covariance matrices of Gaussian states. Given an arbitrary $n$-mode Gaussian state with covariance matrix $\gamma$, there exists a symplectic matrix $S$ such that 
\begin{align}
\gamma=S[\bigoplus_{k=1}^{n}\nu_{k} I_{2}]S^{T},  \label{williamson}
\end{align}
where $S$ is a symplectic matrix, i.e., $S\Omega S^{T}=\Omega$. The $n$ positive quantities $\nu_{k}$ are called the symplectic eigenvalues of $\gamma$, and can also be computed by taking the modulus of the standard eigenspectrum of the matrix $i\Omega \gamma$. 
The symplectic spectrum expresses the fundamental properties of the corresponding Gaussian  quantum state. For example, the uncertainty principle \eqref{uncertainty} is equivalent to 
$\nu_{k}\ge 1$. When $\nu_{k}= 1$ for all $k$, the resulting covariance matrix $\gamma=SS^{T}$ corresponds to a pure Gaussian state~\cite{SSM87:pra,SSM88:pra,MFL11:pra}. 

Furthermore, the symplectic matrix $S$ in Eq.~\eqref{williamson} can be decomposed using the Euler decomposition~\cite{ADMS95:Pramana,B05:pra}. In fact, every $n$-mode symplectic matrix $S$ can be written as 
\begin{align}
S=K[\bigoplus_{k=1}^{n}S(r_{k})]L, \label{Sdecomp}
\end{align}
where $K$ and $L$ are symplectic and orthogonal matrices, and $S(r_{k})=\begin{pmatrix}
e^{-r_{k}} &0\\
0 &e^{r_{k}}
 \end{pmatrix}$ is a set of single-mode squeezing matrices. It is worth mentioning that the symplectic and orthogonal matrices $K$ and $L$ correspond to passive interferometers which can be implemented using a network of beam splitters and phase shifters in quantum optics.   
Combining Eq.~\eqref{williamson} and Eq.~\eqref{Sdecomp}, we obtain that an arbitrary $n$-mode covariance matrix $\gamma$ can be written as 
\begin{align}
\gamma=K[\bigoplus_{k=1}^{n}S(r_{k})]L[\bigoplus_{k=1}^{n}\nu_{k} I_{2}]L^{T}[\bigoplus_{k=1}^{n}S(r_{k})] K^{T}.   \label{williamson2}
\end{align}
Physically, Eq.~\eqref{williamson2} means that every $n$-mode zero-mean Gaussian quantum state can be prepared beginning with $n$ thermal states described by the diagonal covariance matrix $\bm{\nu}=\bigoplus_{k=1}^{n}\nu_{k} I_{2}$,  followed by applying an interferometer $L$, then single-mode squeezers $S(r_{k})$ and finally an interferometer $K$. 

In this section, our main objective is to construct an example of a $2\times 2$  bound entangled Gaussian state that is simple to be prepared in quantum optics. This is done by using Eq.~\eqref{williamson2} with the help of the numerical detection method discussed in Sec.~\ref{detection}. Firstly, we consider the symplectic and orthogonal matrices  $K$ and $L$ in Eq.~\eqref{williamson2}. We rearrange the entries of the matrix $\Omega$ such that we have 
$J:=P^{T}\Omega P=\begin{pmatrix}
0 &I_{4}\\
-I_{4} &0
\end{pmatrix}$ where 
\begin{align*}P=\begin{pmatrix}
     1 &0 &0 &0 &0 &0 &0 &0\\
     0 &0 &0 &0 &1 &0 &0 &0\\
     0 &1 &0 &0 &0 &0 &0 &0\\
     0 &0 &0 &0 &0 &1 &0 &0\\
     0 &0 &1 &0 &0 &0 &0 &0\\
     0 &0 &0 &0 &0 &0 &1 &0\\
     0 &0 &0 &1 &0 &0 &0 &0\\
     0 &0 &0 &0 &0 &0 &0 &1
      \end{pmatrix}.
      \end{align*} 
Let us define $O:=P^{T}KP$. Then it can be shown that  $O$ is orthogonal (i.e., $O^{T}O= I$) and  satisfies $OJO^{T}=J$, which means that the matrix $O$ can be written as 
$O=\begin{pmatrix}
X &Y\\
-Y &X
\end{pmatrix}$,
where $XX^{T}+YY^{T}=I$ and $XY^{T}-YX^{T}=0$. This implies that $Q:=X+iY$ is a unitary  matrix. Therefore, if we choose a unitary matrix $Q$, and  
let  $X := \re(Q)$ and $Y := \im(Q)$, the matrix $O=\begin{pmatrix}
      X &Y\\
     -Y &X
     \end{pmatrix}$ is orthogonal and satisfies $OJO^{T}=J$. As a result, the corresponding symplectic and orthogonal matrix $K$ can be obtained by 
$K=POP^{T}$. We mention that  the class of $4\times 4$ unitary matrices has been parametrized in  Ref.~\cite{D82:jpa}.  After determining $K$ and $L$, then we choose symplectic eigenvalues $\nu_{k}$ and squeezing matrices $S(r_{k})$ in Eq.~\eqref{williamson2} and we obtain a Gaussian state $\gamma$. This  Gaussian state $\gamma$  is then tested by the numerical detection method discussed in Sec.~\ref{detection}. If the obtained Gaussian state $\gamma$ satisfies the PPT criterion~\eqref{PPTcriterion}, but nevertheless is entangled, then we obtain a bound entangled state. Otherwise, we try different parameters until we get a bound entangled state. 
 
Using the idea above, we have successfully obtained a $2\times 2$ bound entangled Gaussian state. The covariance matrix of this Gaussian state is calculated as
\begin{align}
&\gamma=\notag\\
&\begin{pmatrix}
    1.8605   &      0   &-0.7593   &      0    &0.1030   &      0   & 0.6384   &      0\\
         0   & 1.8687   &      0   &-0.3556    &     0   & 0.6854   &      0   & 1.0340\\
   -0.7593   &      0   & 2.3534   &      0    &1.0738   &      0   &-0.7593   &      0\\
         0   &-0.3556   &      0   & 1.9334    &     0   & 0.5029   &      0   &-0.3556\\
    0.1030   &      0   & 1.0738   &      0    &2.4990   &      0   & 0.1030   &      0\\
         0   & 0.6854   &      0   & 0.5029    &     0   & 2.9027   &      0   & 0.6854\\
    0.6384   &      0   &-0.7593   &      0    &0.1030   &      0   & 1.8605   &      0\\
         0   & 1.0340   &      0   &-0.3556    &     0   & 0.6854   &      0   & 1.8687
\end{pmatrix}. \label{robustcov}
\end{align}
\normalsize
We obtain the covariance matrix~\eqref{robustcov} by using the following procedure. First, we choose the symplectic eigenvalues $\nu_{k}\ge 1$ in Eq.~\eqref{williamson2}. Here  we have chosen 
\begin{align}
\bm{\nu}=\bigoplus_{k=1}^{n}\nu_{k} I_{2}=\begin{pmatrix}
1.01 &&&&&&& \mathbf{0}\\
 &1.01\\
 &&1.01\\
 &&&1.01\\
 &&&&3.2\\
 &&&&&3.2\\
 &&&&&&3.2\\
\mathbf{0} &&&&&&&3.2 
\end{pmatrix}. \label{nu}
\end{align}
Second, we choose the symplectic and orthogonal matrix $L$ and the squeezing operators $S(r_{k})$. Here for simple implementation of the resulting state, we have chosen $L=I$ and 
 \begin{align}
 S(r_{1})&=S(r_{3})=\begin{pmatrix}
 1.1 &0\\
 0 &\frac{1}{1.1}
 \end{pmatrix}, \label{squeezer12} \\
 S(r_{2})&=S(r_{4})=\begin{pmatrix}
 \frac{1}{1.1} &0\\
 0 &1.1
 \end{pmatrix}. \label{squeezer34}
 \end{align}
Finally, we choose the symplectic matrix $K$. Using the method described before, we  choose the unitary matrix 
$Q=\begin{pmatrix} 
\frac{\sqrt{2}}{2} &\frac{\sqrt{2}}{4} &-\frac{\sqrt{2}}{4} &\frac{1}{2}\\
0 &\frac{\sqrt{2}}{2} &\frac{\sqrt{2}}{2} &0\\
0 &-\frac{1}{2} &\frac{1}{2} &\frac{\sqrt{2}}{2}\\
-\frac{\sqrt{2}}{2} &\frac{\sqrt{2}}{4} &-\frac{\sqrt{2}}{4} &\frac{1}{2}
\end{pmatrix}$. The resulting symplectic and orthogonal matrix $K$ is then calculated as  
    \begin{align*}
 K&=\begin{pmatrix}
  \frac{\sqrt{2}}{2} &0 &\frac{\sqrt{2}}{4}  &0 &-\frac{\sqrt{2}}{4} &0 &\frac{1}{2}            &0\\
    0 &\frac{\sqrt{2}}{2} &0 &\frac{\sqrt{2}}{4} &0 &-\frac{\sqrt{2}}{4} &0             &\frac{1}{2}\\
0 &0 &\frac{\sqrt{2}}{2} &0 &\frac{\sqrt{2}}{2} &0 &0  &0 \\
  0 &0  &0 &\frac{\sqrt{2}}{2} &0  & \frac{\sqrt{2}}{2} &0  &0 \\
0 &0 &-\frac{1}{2} &0 &\frac{1}{2} &0 &\frac{\sqrt{2}}{2} &0 \\
   0 &0  &0 &-\frac{1}{2} &0 &\frac{1}{2} &0 &\frac{\sqrt{2}}{2} \\
-\frac{\sqrt{2}}{2} &0 &\frac{\sqrt{2}}{4}  &0 &-\frac{\sqrt{2}}{4} &0 &\frac{1}{2}            &0\\
    0 &-\frac{\sqrt{2}}{2} &0 &\frac{\sqrt{2}}{4} &0 &-\frac{\sqrt{2}}{4} &0             &\frac{1}{2}
     \end{pmatrix}. 
 \end{align*}

Substituting the above values of $\bm{\nu}$, $L$, $S(r_{k})$ and $K$ into the decomposition~\eqref{williamson2}, we will obtain the covariance matrix~\eqref{robustcov}.  The physicality of the state~\eqref{robustcov} is guaranteed by our choice of the symplectic eigenvalues $\nu_{k}\ge 1$. Next we show that it is a bound entangled state. It is found that $\text{min}\; \text{eig}(\gamma+i\tilde{\Omega})=0.0840>0$, thus the covariance matrix~\eqref{robustcov} satisfies the PPT criterion~\eqref{PPTcriterion} and is not distillable. On the other hand, we find by numerical computation that the set of strict inequalities \eqref{constraint4}-\eqref{constraint6} with $\epsilon=10^{-8}$ is infeasible given the covariance matrix~\eqref{robustcov}. Hence, it is an entangled state. In conclusion, the $2\times 2$ Gaussian state \eqref{robustcov} satisfies the PPT criterion~\eqref{PPTcriterion}, but is an entangled state.  So it is a bound entangled state. 

\begin{figure}[htbp]
\begin{center}
\includegraphics[height=5cm]{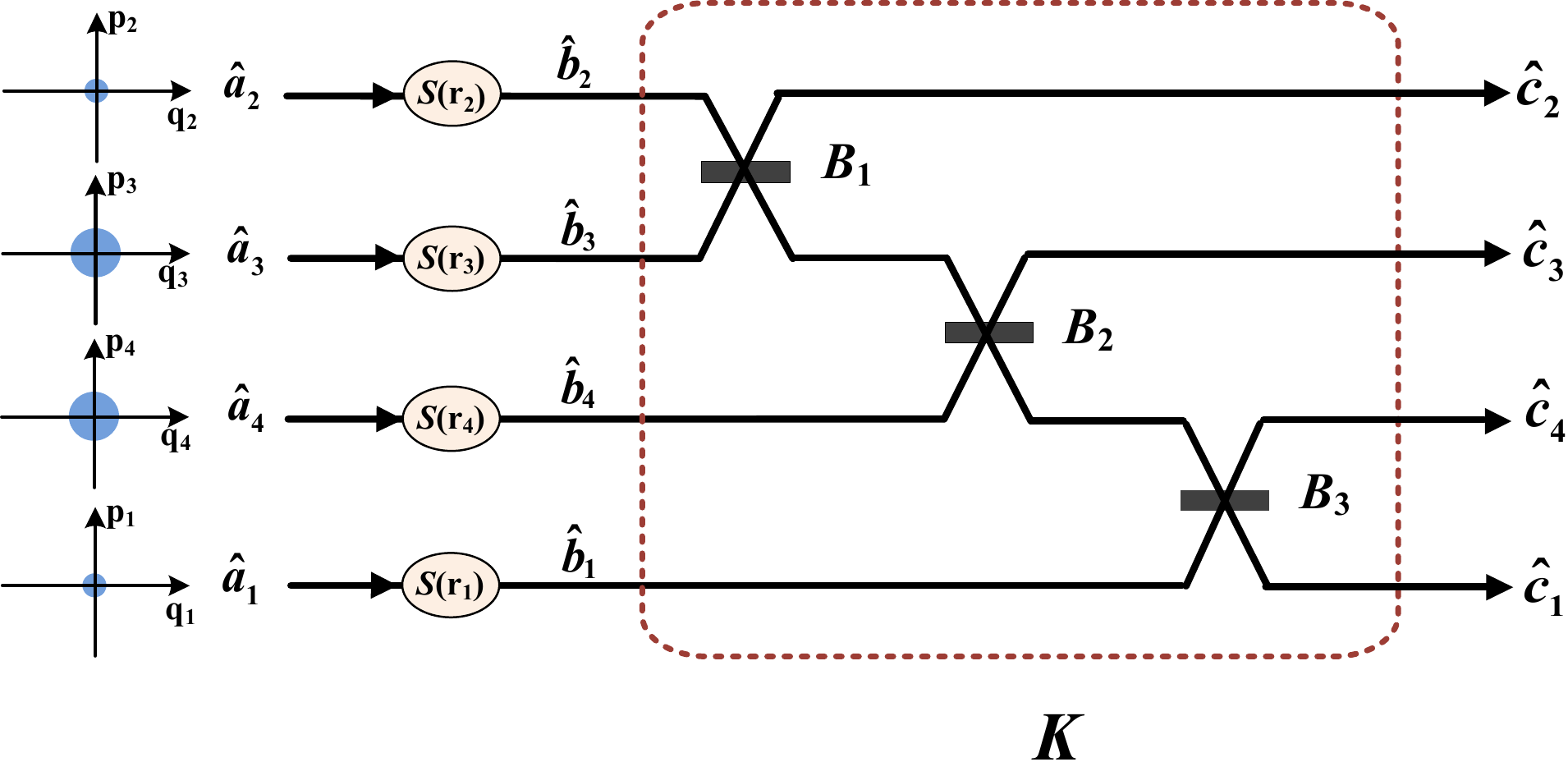}
\caption{Diagram for preparation of the bound entangled Gaussian state~\eqref{robustcov} in quantum optics. The initial inputs $\hat{a}_{1}$,  $\hat{a}_{2}$, $\hat{a}_{3}$ and $\hat{a}_{4}$ are in thermal states described by the covariance matrix $\nu$ shown in Eq.~\eqref{nu}. $B_{1}$, $B_{2}$ and $B_{3}$ are balanced beam splitters and realize the corresponding unitary transformations \eqref{beam1} and \eqref{beam2};   
$S(r_{1})$, $\cdots$, $S(r_{4})$ are a set of single-mode squeezers as described by Eq.~\eqref{squeezer12} and \eqref{squeezer34}. The output Gaussian state ($\hat{c}_{1},\cdots,\hat{c}_{4}$) has the covariance matrix \eqref{robustcov}, and is a $2\times 2$ bound entangled state with respect to the bipartite splitting $\{\{\hat{c}_{1},\hat{c}_{2}\},\;\{\hat{c}_{3},\hat{c}_{4}\}\}$.}
\label{all_realize}
\end{center}
\end{figure}

Now we show how to construct an optical system to generate the bound entangled Gaussian state~\eqref{robustcov}. We note that the symplectic map $K$ corresponds to the following linear unitary transformation on the annihilation operators
\begin{align*}
\begin{pmatrix}
\hat{c}_{1}\\
\hat{c}_{2}\\
\hat{c}_{3}\\
\hat{c}_{4}
\end{pmatrix}=\begin{pmatrix}
\frac{\sqrt{2}}{2} &\frac{\sqrt{2}}{4}  &-\frac{\sqrt{2}}{4}  &\frac{1}{2}\\
0   &\frac{\sqrt{2}}{2} &\frac{\sqrt{2}}{2}  &0\\
0 &-\frac{1}{2}  &\frac{1}{2} &\frac{\sqrt{2}}{2}\\
-\frac{\sqrt{2}}{2} &\frac{\sqrt{2}}{4} &-\frac{\sqrt{2}}{4}  &\frac{1}{2}
\end{pmatrix}\begin{pmatrix}
\hat{b}_{1}\\
\hat{b}_{2}\\
\hat{b}_{3}\\
\hat{b}_{4}
\end{pmatrix}.
\end{align*}
Here $\hat{b}_{1}-\hat{b}_{4}$ and $\hat{c}_{1}-\hat{c}_{4}$ are annihilation operators; see Fig.~\ref{all_realize} for details.  Furthermore, using the result developed in Ref.~\cite{RZBB94:prl,CHMKW16:opt}, this unitary transformation can be realized as a network of three beam splitters  as follows 
\begin{align}
&\begin{pmatrix}
\frac{\sqrt{2}}{2} &\frac{\sqrt{2}}{4}  &-\frac{\sqrt{2}}{4}  &\frac{1}{2}\\
0   &\frac{\sqrt{2}}{2} &\frac{\sqrt{2}}{2}  &0\\
0 &-\frac{1}{2}  &\frac{1}{2} &\frac{\sqrt{2}}{2}\\
-\frac{\sqrt{2}}{2} &\frac{\sqrt{2}}{4} &-\frac{\sqrt{2}}{4}  &\frac{1}{2}
\end{pmatrix} =B_{3}B_{2}B_{1}, \label{Kdecomposition}
\end{align}
where $B_{1}-B_{3}$, representing beam-splitter transformations, are given by 
\begin{align}
B_{1}&=\begin{pmatrix}
1 &0 &0 &0\\
0 &\frac{\sqrt{2}}{2} &\frac{\sqrt{2}}{2} &0\\
0 &-\frac{\sqrt{2}}{2} &\frac{\sqrt{2}}{2} &0\\
0 &0 &0 &1
\end{pmatrix},
B_{2}=\begin{pmatrix}
1 &0 &0 &0\\
0 &1 &0 &0\\
0 &0 &\frac{\sqrt{2}}{2} &\frac{\sqrt{2}}{2}\\
0 &0 &-\frac{\sqrt{2}}{2} &\frac{\sqrt{2}}{2}
\end{pmatrix},\label{beam1}\\
B_{3}&=\begin{pmatrix}
\frac{\sqrt{2}}{2} &0 &0 &\frac{\sqrt{2}}{2}\\
0 &1 &0 &0\\
0 &0 &1 &0\\
-\frac{\sqrt{2}}{2} &0 &0 &\frac{\sqrt{2}}{2}
\end{pmatrix}. \label{beam2}
       \end{align}
Hence the symplectic map $K$ is implemented by a network of beam splitters  as described in the dotted box in Fig.~\ref{all_realize}.  $B_{1}$, $B_{2}$ and $B_{3}$ are balanced beam splitters ($50:50$). According to Eq.~\eqref{williamson2}, the output Gaussian state ($\hat{c}_{1},\cdots,\hat{c}_{4}$) generated by the optical system described in Fig.~\ref{all_realize} has the covariance matrix \eqref{robustcov}, and is a bound entangled state with respect to the bipartite splitting $\{\{\hat{c}_{1},\hat{c}_{2}\},\;\{\hat{c}_{3},\hat{c}_{4}\}\}$.

The bound entangled Gaussian state  \eqref{robustcov}  has some nice characteristics as follows that make it simple enough to be producible in quantum optics. 
\begin{enumerate}
\item $L=I$ in Eq.~\eqref{williamson2}. It means that we do not need to implement another beam splitter network before the single-mode squeezers $S(r_{1})$, $\cdots$, $S(r_{4})$.
\item The optical system only consists of three beam splitters $B_{1}$, $B_{2}$ and $B_{3}$. It happens that this is  the minimum number of beam splitters to prepare a $2\times 2$ bound entangled Gaussian state. Besides, all the beam splitters $B_{1}$, $B_{2}$ and $B_{3}$ are balanced ($50:50$), and hence they can be easily implemented experimentally.
\item As shown in Eq.~\eqref{nu}, all the symplectic eigenvalues $\nu_{k}$ are strictly larger than $1$. In other words, the initial inputs $\hat{a}_{1}$ - $\hat{a}_{4}$ are all in thermal states. This is clearly realistic since in
actual implementations we cannot prepare precisely vacuum inputs due to the presence of noise  and imperfections.  For comparison, the preparation of the bound entangled state \eqref{bound_state_example} requires two vacuum inputs as illustrated in Ref.~\cite{MWJZ19:pra}. 
\item The state \eqref{robustcov} is fairly robust for
experimental verification. Numerical simulations show that small imperfections in the implementation (such as input fields and squeezers) can still generate bound entanglement (though not the original state). An important reason is that the state \eqref{robustcov} has strictly positive partial transpose, i.e., $ \gamma+i\tilde{\Omega}>0$. Hence small imperfections in experimental implementations still lead to  a PPT state. 
\end{enumerate}
Because of the above characteristics, the Gaussian state \eqref{robustcov} may serve as a good candidate  for a realistic experimental verification of bound entanglement.  
\section{Conclusion}
In this paper, we have shown that the separability problem of Gaussian quantum states can be cast as  an equivalent  problem of determining feasibility of a set of linear matrix inequalities. Thus, Gaussian entanglement can be directly detected by checking the feasibility of the corresponding linear matrix inequalities. We have applied this method to the identification of bound entangled Gaussian states. By choosing some parameters in the decomposition of a covariance matrix, we can find bound entangled states that are simple enough to be producible in quantum optics. We have provided an optical scheme for generating a particular bound entangled Gaussian state in quantum optics. We have highlighted some characteristics of the bound entangled Gaussian state which may make it a good candidate  for a realistic experimental verification of bound entanglement.  In future work, it would be interesting to investigate potential applications of bound entangled Gaussian states to quantum engineering areas such as multiparty quantum communication \cite{AH09:pra,JZWZXP12:prl}, and quantum metrology~\cite{NLKA18:pra,LYLW20:jpa}. 

 \section{Acknowledgment}
We thank Jing Zhang and Xiaojun Jia for
helpful discussions. This work was supported in part by the National Natural Science Foundation of China (NSFC) under Grant Nos. 61803389, 61873162 and 61973317,  the Shanghai Pujiang Program (Grant No. 18PJ1405500) and the Hunan Provincial Natural
Science Foundation of China (Grant No. 2017JJ2272).

\end{document}